\documentclass[%
aip,
 amsmath,amssymb,
 reprint,%
]{revtex4-1}

\usepackage{graphicx}
\usepackage{dcolumn}
\usepackage{bm}
\usepackage{cleveref}
\usepackage[normalem]{ulem}
\usepackage{xcolor}

\usepackage[utf8]{inputenc}
\usepackage[T1]{fontenc}
\usepackage{mathptmx}
\usepackage{etoolbox}
\usepackage{bibentry}
\usepackage{rotating}
\usepackage{marginnote}
\usepackage{hyperref}
\newcommand{\verticalmarginnote}[1]{\marginnote{\hspace{27em}\begin{turn}{90}\begin{minipage}{20cm}\raggedright\small #1\end{minipage}\end{turn}}}

\makeatletter
\def\@email#1#2{%
 \endgroup
 \patchcmd{\titleblock@produce}
  {\frontmatter@RRAPformat}
  {\frontmatter@RRAPformat{\produce@RRAP{*#1\href{mailto:#2}{#2}}}\frontmatter@RRAPformat}
  {}{}
}%
\makeatother
\newcommand{\replace}[2]{{#2}}

\begin{document}

\preprint{AIP/123-QED}

\title[Brute-force nucleation rates of hard spheres]{Brute-force nucleation rates of hard spheres compared with rare-event methods and classical nucleation theory}
\author{Willem Gispen}
\author{Marjolein Dijkstra}%
 \email{m.dijkstra@uu.nl}
\affiliation{ 
Soft Condensed Matter \& Biophysics, Debye Institute for Nanomaterials Science, Utrecht University, Princetonplein 1, 3584 CC Utrecht, Netherlands 
}%


\date{\today}

\begin{abstract}
We determine the nucleation rates of hard spheres using brute-force molecular dynamics simulations.
We overcome nucleation barriers of up to $28 ~k_B T$, leading to a rigorous test of nucleation rates obtained from rare-event methods and classical nucleation theory. Our brute-force nucleation rates show excellent agreement with umbrella sampling simulations by Filion \textit{et al.}\ [J. Chem. Phys. \textbf{133},
244115 (2010)]
and seeding simulations by Espinosa \textit{et al.}\ [J. Chem. Phys. \textbf{144},
034501 (2016)].

\end{abstract}

\verticalmarginnote{This article may be downloaded for personal use only. Any other use requires prior permission of the author and AIP Publishing.\\This article appeared in the Journal of Chemical Physics and may be found at \href{https://doi.org/10.1063/5.0165159}{https://doi.org/10.1063/5.0165159}}

\maketitle



The formation of crystals through nucleation occurs in natural phenomena such as ice formation in clouds and crystallization of the earth's inner core,\cite{sun_two-step_2022} as well as industrial processes such as self-assembly of soft materials and pharmaceuticals. Therefore, the calculation of absolute nucleation rates of crystallization with computer simulations is useful for understanding, predicting, and controlling these processes. It is of course essential to validate these calculations in controlled experiments. Colloidal hard spheres are a useful model system for this purpose, as their crystallization occurs on time and length scales that are more easily accessible than their molecular or atomic counterparts.

However, in experimental realizations of colloidal hard spheres, crystal nuclei form up to 10 billion times faster than predicted by simulations.\cite{auer_prediction_2001} 
To understand \replace{the}{this} discrepancy, previous work \replace{}{-- reviewed recently in Ref. \citenum{royall_colloidal_2023} --} has mostly focused on the modeling of experimental effects or interpretation of experimental measurements\replace{\cite{royall_colloidal_2023}}{} such as the role of sedimentation,\cite{russo_interplay_2013} hydrodynamics,\cite{fiorucci_effect_2020} or heterogeneous nucleation.\cite{respinosa_heterogeneous_2019,wohler_hard_2022}
In contrast, there are fewer direct tests of the internal consistency of theoretical predictions. The reason is simple: direct simulation of crystal nucleation becomes exponentially more expensive as one lowers the supersaturation.

Therefore, to predict the nucleation rate in this regime, previous work has heavily relied on rare-event methods such as umbrella sampling\cite{auer_prediction_2001,filion_crystal_2010} and forward-flux sampling.\cite{filion_crystal_2010} Another line of approach is to use a combination of classical nucleation theory and simulations.\cite{espinosa_seeding_2016} We note that the regime of largest discrepancy between simulation and experiment is precisely in the low supersaturation regime, where rare-event methods are necessary. In contrast, brute force simulations at high supersaturation do not seem to have such a large discrepancy with experiments. This raises the possibility that in fact rare-event methods could be unreliable in the low supersaturation regime.

A question that is intimately related is the question whether nucleation can be described accurately with a one-dimensional order parameter. In classical nucleation theory, the nucleus size, i.e.\ the number of particles in the crystal nucleus, is the only order parameter. In fact, in rare-event methods, the nucleus size is computed to analyze and bias the simulations to observe nucleation. On the other hand, there are more and more doubts that a one-dimensional order parameter is appropriate for nucleation\replace{.}{, e.g.\ in the case of two-step nucleation.} For example, committor analyses show that the nucleus size is not a perfect reaction coordinate,
and that the bond orientational order of the crystal nucleus should also be taken into account.\cite{moroni_interplay_2005}
As rare-event methods have been shown to depend on a good reaction coordinate,\cite{jungblut_reaction_2015,blow_seven_2021} using the nucleus size as the only order parameter could lead to severe errors in calculations of the nucleation rate.

In this Note, we compute the nucleation rates of hard spheres using brute-force simulations in a low supersaturation regime. Next, we compare the nucleation rates with rare-event methods and classical nucleation theory. In this way, we perform a rigorous test of these methods.

We simulate a system of nearly hard spheres interacting with a Weeks-Chandler-Andersen (WCA) potential, which is widely used to mimic hard spheres in molecular dynamics simulations. The phase behavior of the WCA system maps very well on hard spheres if an effective hard-sphere diameter is defined.\cite{filion_simulation_2011} The effective hard-sphere diameter \replace{$\sigma_{\textrm{eff}}$}{$\sigma_{\textrm{eff}}=1.097\sigma$} is defined such that the freezing density of the WCA system\cite{filion_simulation_2011} maps onto the freezing density of hard spheres.\cite{frenkel2001understanding} This mapping has been shown to accurately capture the \replace{}{supersaturation and} nucleation rates of hard spheres.\cite{filion_simulation_2011} To be precise, we perform molecular dynamics (MD) simulations in the canonical ($NVT$) ensemble with $N=2\times10^4$ particles at a temperature $k_B T/\epsilon = 0.025$ and vary the effective packing fraction $\eta_{\mathrm{eff}}=\pi N \sigma_{\mathrm{eff}}^3 / 6V$ \replace{}{from $0.5283$ to $0.5366$. Our system size is sufficiently large to avoid self-interaction of the critical nucleus, which contains at most $100$ particles under these conditions. \cite{filion_crystal_2010}} The equations of motion are integrated with a Nose-Hoover thermostat implemented in the LAMMPS molecular dynamics code, \cite{plimpton_fast_1995,shinoda_rapid_2004}\replace{ and}{} a timestep $\Delta t = 0.001\sqrt{m\sigma^2 / k_B T}$\replace{.}{ where $m$ is the particle mass, and a thermostat relaxation time of $100$ timesteps}. We choose the canonical ensemble and this moderate system size to optimize the efficiency of the simulations.

To calculate a nucleation rate with brute force simulations, we \replace{assume}{use the standard assumption\cite{filion_crystal_2010,filion_simulation_2011,blow_seven_2021,wohler_hard_2022}} that the nucleation times are distributed according to an exponential distribution\replace{.}{ and verified this assumption using the survival function.} \replace{To monitor the system, we use the pressure, which sharply decreases during crystallization. Once that happens, we immediately stop the simulation.}{ During crystallization at constant volume, the pressure sharply decreases. We use this as a marker to identify a spontaneous nucleation event, record the time until crystallization $t_i$, and stop the simulation. If a sample $i$ does not crystallize, we just record the final simulation time $t_i$. The total simulation time $t=\sum_i t_i$ is then the sum of all simulation times $t_i$, whether they crystallized or not.} Given a number of observed spontaneous nucleation events \replace{$n$}{$\ell$} in a total simulation time $t$ and volume $V$, the nucleation rate density is estimated as
\begin{equation*}
 \label{eq:rate}
    J = \ell / (V t).
\end{equation*}
Note that this expression is appropriate for a \replace{truncated}{censored} exponential distribution\replace{.}{, i.e.\ our case where not all samples have crystallized.\cite{wohler_hard_2022,deemer_estimation_1955}} For low supersaturation, the nucleation rate is low, and consequently the number of observed nucleation events is limited. \replace{}{In our case, we are mainly interested in the order of magnitude of the nucleation rate, and therefore only a few observations are needed to obtain a reasonable estimate.} We quantify the uncertainty in the nucleation rate estimated from \replace{$n$}{$\ell$} nucleation events using a chi-squared distribution with \replace{$2n$}{$2\ell$} degrees of freedom. \replace{}{To be precise, from the 2.5th and 97.5th percentiles $P_{2.5}$ and $P_{97.5}$ of this distribution, we approximate the $95\%$ confidence interval of the nucleation rate $J$ as $J P_{2.5}/2\ell < J < J P_{97.5}/2\ell$.\cite{deemer_estimation_1955}} \replace{In our case, we are mainly interested in the order of magnitude of the nucleation rate, and therefore only a few observations are needed to obtain a reasonable estimate.}{}

\begin{figure}[t]
    \centering
    \includegraphics[width=\linewidth]{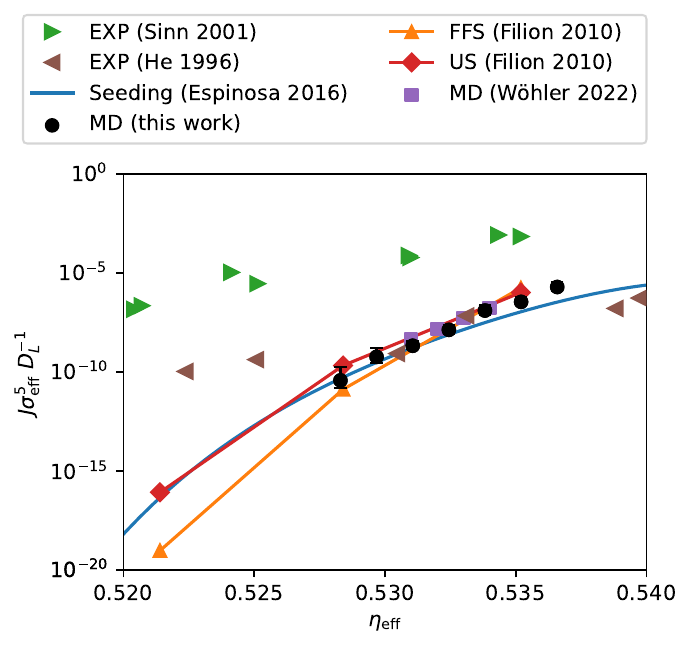}
    \caption{Nucleation rates $J \sigma_{\mathrm{eff}}^5 D_L^{-1}$ of hard spheres as a function of effective packing fraction $\eta_{\mathrm{eff}}$, where $D_L$ denotes the long-time self-diffusion coefficient. The brute-force nucleation rates computed in this work are shown in black and compared to \replace{}{previous results from molecular dynamics (MD),\cite{wohler_hard_2022}} forward flux sampling (FFS),\cite{filion_crystal_2010} umbrella sampling (US),\cite{filion_crystal_2010} seeding,\replace{\cite{sanchez-burgos_fcc_2021}}{\cite{espinosa_seeding_2016}} and experiments (EXP).\cite{he_dynamics_1996,sinn_solidification_2001}
    }
    \label{fig:freezing-rates}
\end{figure}

\begin{table}[]
    \centering
    \begin{tabular}{c|c|c|c|c|c|c}
   $\eta_{\mathrm{eff}}$ & $N \sigma^3/V$ & $\ell$ & $L$ & $J \sigma_{\mathrm{eff}}^5 D_L^{-1}$ & $P_{2.5}/2\ell$ & $P_{97.5}/2\ell$\\
       \hline
        0.5283 & 0.765 & 3 & 284 & $2 \times 10^{ -11 }$ & 0.4 & 4.9 \\
        0.5297 & 0.767 & 6 & 32 & $4 \times 10^{ -10 }$ & 0.5 & 2.7 \\
        0.5311 & 0.769 & 17 & 32 & $1 \times 10^{ -9 }$ & 0.7 & 1.7 \\
        0.5324 & 0.771 & 16 & 16 & $8 \times 10^{ -9 }$ & 0.6 & 1.8 \\
        0.5338 & 0.773 & 16 & 16 & $8 \times 10^{ -8 }$ & 0.6 & 1.8 \\
        0.5352 & 0.775 & 17 & 17 & $2 \times 10^{ -7 }$ & 0.7 & 1.7 \\
        0.5366 & 0.777 & 16 & 16 & $1 \times 10^{ -6 }$ & 0.6 & 1.8 \\
    \end{tabular}
    \caption{\replace{}{Nucleation rates $J \sigma_{\mathrm{eff}}^5 D_L^{-1}$ of hard spheres as a function of effective packing fraction $\eta_{\mathrm{eff}}=\pi N \sigma_{\mathrm{eff}}^3 / 6V$. The nucleation rates are estimated from $\ell$ nucleation events observed in a total of $L$ simulations. The $95\%$ confidence interval of the nucleation rate $J$ is given by $J P_{2.5}/2\ell < J < J P_{97.5}/2\ell$.}}
    \label{tab:freezing-rates}
\end{table}

In \Cref{fig:freezing-rates}, we show the nucleation rates of hard spheres as a function of effective packing fraction $\eta_{\mathrm{eff}}$.
Our own brute-force molecular dynamics results are denoted in black\replace{ with error bars representing approximate $95\%$ confidence intervals.}{. The error bars are approximately half an order of magnitude at most and represent the $95\%$ confidence intervals, which are also given in \Cref{tab:freezing-rates}. We normalized the nucleation rate with the long-time self-diffusion coefficient $D_L$, which we computed using independent simulations of the fluid.} In \replace{the same figure, we}{\Cref{fig:freezing-rates}, we also} show the nucleation rates obtained \replace{}{previously} with 
\replace{}{molecular dynamics,\cite{wohler_hard_2022} }
forward flux sampling,\cite{filion_crystal_2010} umbrella sampling, \cite{filion_crystal_2010} seeding,\cite{espinosa_seeding_2016} and experiments using nearly density-matched \cite{he_dynamics_1996} and non-density matched \cite{sinn_solidification_2001} systems. \replace{}{These experiments used light scattering to detect crystallization of colloidal poly(methyl methacrylate) spheres of $0.4$ \textmu m and $0.9$ \textmu m in diameter, respectively.}

Note that our brute-force simulations extend to $\eta_{\mathrm{eff}}=0.5283$, where the nucleation barrier is more than $27.5 k_B T$.\cite{filion_crystal_2010} This corresponds to a nucleation rate two to three orders of magnitude lower than previous brute-force simulations.\cite{filion_crystal_2010,wohler_hard_2022,filion_simulation_2011} We would also like to mention that our brute-force simulations cover the nucleation rate regime that is accessible to experiments.\cite{royall_colloidal_2023}

Our brute-force results show excellent agreement with umbrella sampling\cite{filion_crystal_2010} and seeding.\cite{espinosa_seeding_2016} Forward flux sampling \replace{}{slightly} underestimates the nucleation rate, which is in line with previous results.\cite{haji-akbari_forward-flux_2018} \replace{Overall, t}{T}he largest difference within the simulation results occurs between umbrella sampling and forward-flux sampling at $\eta_{\mathrm{eff}}=0.521$, where the difference is around three orders of magnitude. Overall, these results support the use of \replace{umbrella sampling and seeding}{rare-event methods and classical nucleation theory} for the prediction of nucleation rates.

The agreement with the classical nucleation theory (CNT) prediction \replace{}{-- based on a single order parameter -- } via seeding\cite{espinosa_seeding_2016} is especially remarkable, as CNT has often been argued to give wildly incorrect predictions. \cite{blow_seven_2021} \replace{}{Of course, the excellent agreement we observe for hard spheres does not necessarily translate directly to other systems, where the assumptions of CNT may be less appropriate.} As is well-known, the CNT prediction is very sensitive to the employed values of the interfacial free energy and the supersaturation. Our results show that it is certainly possible to obtain accurate predictions from CNT using seeding,\cite{espinosa_seeding_2016} as long as these two terms are carefully estimated.

However, the discrepancy with experiments remains. 
The potential issues of rare-event methods and classical nucleation theory\cite{blow_seven_2021} result in errors that are small in comparison to the difference with experiments. Therefore, these are not likely to be the origin of the discrepancy. \replace{}{Although the other potential explanations discussed in Ref. \citenum{royall_colloidal_2023} cannot be fully dismissed, the discrepancy may also stem from the determination of packing fraction in the experiments.\cite{paulin_observation_1990,royall_colloidal_2023}}

\begin{acknowledgments}
M.D. and W.G. acknowledge funding from the European Research Council (ERC) under the European Union’s
Horizon 2020 research and innovation programme (Grant
agreement No. ERC-2019-ADG 884902 SoftML).
\end{acknowledgments}

\section*{Data Availability Statement}
The data that supports the findings of this study are available within the article.

\bibliography{refs}

\end{document}